\documentclass{article}
\usepackage[utf8]{inputenc}
\usepackage{authblk}
\usepackage{changepage} 
\usepackage{natbib}
\usepackage{graphicx}
\usepackage{amssymb}

\usepackage{natbib}
\setcitestyle{authoryear,open={(},close={)}}

\providecommand{\keywords}[1]
{
  \small	
  \noindent
  \textbf{Keywords:} #1
}

\author{Geoff Keeling, Google Research \\ gkeeling [at] google.com}
\title{Algorithmic Bias, Generalist Models, \\ and Clinical Medicine}
\date{May 2023}

\begin{document}

\maketitle

\begin{abstract}
  The technical landscape of clinical machine learning is shifting in ways that destabilize pervasive assumptions about the nature and causes of algorithmic bias. On one hand, the dominant paradigm in clinical machine learning is narrow in the sense that models are trained on biomedical datasets for particular clinical tasks such as diagnosis and treatment recommendation. On the other hand, the emerging paradigm is generalist in the sense that general-purpose language models such as Google’s BERT and PaLM are increasingly being adapted for clinical use cases via prompting or fine-tuning on biomedical datasets. Many of these next-generation models provide substantial performance gains over prior clinical models, but at the same time introduce novel kinds of algorithmic bias and complicate the explanatory relationship between algorithmic biases and biases in training data. This paper articulates how and in what respects biases in generalist models differ from biases in prior clinical models, and draws out practical recommendations for algorithmic bias mitigation.
\end{abstract}


\section{Introduction}

Statistical models, including deep neural networks trained via machine learning, are used in medicine for risk assessment, diagnosis, and treatment recommendation, among other clinical use cases.\footnote{For an introduction to machine learning in medicine see \citet{rajkomar2019machine}.} These models can exhibit biases. For example, a model that classifies dermatologic conditions based on images of skin lesions may be less accurate for images featuring darker skin tones compared to lighter skin tones \citep{groh2021evaluating, daneshjou2022disparities}. Typically, such biases arise because, \textit{mutatis mutandis}, darker skin tones are underrepresented in the model’s training data such that the model either fails to register predictively salient statistical patterns that hold for the underrepresented group, or else generalizes a pattern that holds for members of the group sampled but does not hold for that group at the population level.\footnote{Performance biases can also arise, among other reasons, from misrepresentative training data such as datasets that employ proxy variables or data labels that systematically distort the circumstances of a disadvantaged group \citep{obermeyer2019dissecting, jiang2020identifying}. See Section 2.3 for discussion.} These biases are morally significant because at scale biased models threaten to sustain or exacerbate existing inequities in healthcare \citep{panch2019artificial, grote2022algorithmic}.

What is central to the conception of algorithmic bias above is the \textit{ontic assumption} that algorithmic biases are, roughly, measurable performance disparities across demographic groups; and the \textit{explanatory assumption} that such disparities admit explanation in terms of data biases like underrepresentation or misrepresentation of groups in training data. These assumptions are pervasive. Take the US Food and Drug Administration’s (FDA) 2021 report on Software as a Medical Device (SaMD) systems that use machine learning: ‘[b]ecause [such] systems are [...] trained using data from historical datasets, they are vulnerable to bias - and prone to mirroring biases present in the data’ \citep[p.5]{food2021artificial}. This picture is echoed in the FDA’s 2021 principles on ‘Good Machine Learning Practice for Medical Device Development,’ released alongside the UK Medicines and Healthcare Products Regulatory Agency and Health Canada. According to these principles, developers should ensure that ‘datasets are representative of the target population [to] manage any bias [and] promote appropriate and generalizable performance across the intended population’ \citep[p.2]{us2021good}. Likewise, in the scientific literature, \citet[p.1]{norori2021addressing} state that certain subpopulations ‘[are] absent or misrepresented in existing biomedical datasets [such that clinical models are] prone to reinforcing bias’ \citep[c.f.][]{obermeyer2019dissecting, challen2019artificial, cirillo2020sex}.

This paper argues that the technical landscape of clinical machine learning is shifting in ways that destabilize and reframe these pervasive assumptions about the nature and causes of algorithmic bias. On the one hand, the dominant paradigm in clinical machine learning is \textit{narrow} in the sense that models are trained on biomedical datasets for particular clinical tasks such as diagnosis and treatment recommendation. On the other hand, the emerging paradigm is \textit{generalist} in the sense that general-purpose language models such as Google’s BERT \citep{devlin2018bert} and Meta’s OPT \citep{zhang2022opt} are increasingly being adapted for clinical use cases via fine-tuning on biomedical datasets \citep[pp.54-58]{bommasani2021opportunities}. Indeed, BERT has already been adapted for a range of clinical functions including BioBERT for biomedical text mining \citep{lee2020biobert}, G-BERT for treatment recommendation \citep{shang2019pre}, Med-BERT for disease prediction \citep{rasmy2021med}, and ClinicalBERT for predicting hospital readmission \citep{huang2019clinicalbert}. Many of these next-generation models provide substantial performance gains over prior clinical models, but at the same time introduce novel kinds of algorithmic bias, and complicate the explanatory relationship between demographic performance disparities and biases in training data. This paper articulates how and in what respects biases in generalist models differ from biases in previous clinical models, and draws out practical recommendations for algorithmic bias mitigation in SaMD technologies built using generalist language models.

The paper proceeds as follows. Section 2 introduces the narrow paradigm alongside the ontic and explanatory assumptions about algorithmic bias that it gave rise to. Section 3 characterizes the emerging generalist paradigm and shows how biases in this paradigm challenge the prevailing assumptions. Section 4 explores bias mitigation strategies for generalist models. Section 5 concludes.

\section{The Narrow Paradigm}

\subsection{Narrow Models and Supervised Learning}

The core idea of the narrow paradigm is that machine learning models are trained with a particular clinical task in mind. An example of a clinical task is predicting whether a nodule depicted in a computed tomography (CT) image is malignant or benign \citep{uthoff2019differentiation}. Another is predicting based on features of a patient’s medical history whether or not that patient will be readmitted to hospital within 30 days if discharged \citep{hasan2010hospital}.

The dominant approach to machine learning within the narrow paradigm is called \textit{supervised learning}. Here a model is trained to predict an outcome variable based on one or more input features via a large number of examples. So given a dataset comprising a series of CT images each depicting a nodule correctly labeled as ‘benign’ or ‘malignant,’ a model can learn which features of the images predict malignancy through a trial-and-error process where the model parameters are iteratively updated to correct for erroneous predictions \citep{uthoff2019differentiation}. Supervised learning accounts for much of the recent success in clinical machine learning. This includes image classification models that rival the performance of clinicians at dermatologic diagnosis \citep{tschandl2019expert}, and similarly performant models for radiological tasks such as image-based diagnosis, segmentation, and identification \citep{kelly2022radiology}. 

Sometimes supervised learning is supplemented with a technique called transfer learning \citep{weiss2016survey}. Standardly, training data reflects the kind of inputs and outputs that the model will operate with in practice. Hence a model for identifying malignant nodules is typically trained on labeled images of malignant and benign nodules. But in some domains - including medicine - training data is expensive to collate because experts are required to label the data and the quantity of data needed is substantial \citep{rahimi2021addressing}. Transfer learning offers a partial solution. Here the model is pre-trained on a generic task and then fine-tuned for a specific task. The idea is that statistical associations learned in the generic task are applicable in (and thus transfer to) the domain-specific task. This reduces the amount of clinical data that is required to train the model. Hence a model for identifying malignant nodules might first be trained to classify images on a generic image dataset such as ImageNet \citep{deng2009imagenet}, and subsequently be fine-tuned to detect malignant nodules by holding fixed the model weights in the main body of the neural network that account for lower level features such as edges and corners, and re-training the top layers to account for higher level clinically significant features using the nodule image dataset \citep[c.f.][]{chowdhury2021applying}. 

Importantly: Whether or not transfer learning is used the background assumption is that each model is trained to perform a particular clinical task. Pervasive assumptions about the nature and causes of algorithmic bias in medicine have developed on the backdrop of this narrow paradigm. The next two sections articulate and make precise these assumptions.

\subsection{What are Biases?}

On first approximation the ontic assumption holds that algorithmic biases admit analysis in terms of measurable performance disparities across demographic groups. So one respect in which a model for identifying the presence or absence of a particular dermatologic condition based on images of skin lesions might be biased is if the area under the receiver operating characteristic curve (AUC-ROC) is lower for images with Fitzpatrick skin types I-II compared to images with Fitzpatrick skin types V-VI \citep{daneshjou2022disparities}. Here AUC-ROC is a metric that measures the degree to which a binary classifier is able to distinguish between the two classes (in this case the presence or absence of the relevant dermatologic condition). The implication of different AUC-ROC scores for Fitzpatrick skin types I-II and V-VI is that the model is better able to identify the relevant condition for White patients compared to Black patients, as race and Fitzpatrick skin type are correlated, albeit imperfectly \citep{ware2020racial}. Performance disparities across demographic groups like the one described is typically what is at issue when clinical models are claimed to exhibit algorithmic bias \citep{hellstrom2020bias}.

Whether a model exhibits performance disparities across demographic groups is not always clear. Reasonable people can disagree about what counts as good performance. For example, individuals with different attitudes to risk may disagree about the comparative importance of precision and recall in a screening task. Here precision is the fraction of actually positive cases among those classified as positive and recall is the fraction of actually positive cases that end up classified as positive. Risk-neutral people who are indifferent between errors in which actually negative cases are classified as positive and errors in which actually positive cases are classified as negative may advocate for the $F_1$ score as a performance metric, i.e. the harmonic mean of precision and recall. In contrast, risk-averse people who prefer actually negative cases to test positive than for actually positive cases to test negative may advocate for an $F_{\beta}$ score, which is a weighted harmonic mean where recall has $\beta$ times the weight of precision for some $\beta$ $>$ 0. Hence even if algorithmic bias is understood in the sense of measurable demographic performance disparities, there may be disagreements about which models are biased given the lack of consensus on how to measure performance \citep[c.f.][]{dieterich2016compas}.\footnote{The problem is underscored by the fact that absent equal base rates across subpopulations or perfect predictive performance binary classifiers cannot equalize precision and false positive/negative rates \citep{chouldechova2017fair}. See \citet{kleinberg2016inherent} for an analogous result for continuous risk scores. For discussion of the significance of the fairness impossibility theorems for healthcare see \citet{grote2022algorithmic} and \citet{grote2022enabling}.}

\subsection{How are Biases Explained?}

The explanatory assumption holds that algorithmic biases are typically explained by biases in training data. Performance disparities obtain at the model level because the model’s training data underrepresents or misrepresents particular demographic groups. To use the FDA’s term, algorithmic biases are ‘prone to mirroring biases present in the data’ \citet[][p.5, emphasis mine]{food2021artificial}. The relevant sense of \textit{mirroring} is that performance disparities which disadvantage group \textit{G} are explained by underrepresentation or misrepresentation of \textit{G} in the training data.\footnote{The implication here is not that all performance biases are explained by biases in training data, as biases can arise at every stage of the machine learning pipeline \citep{rajkomar2019machine}. Rather, the claim is that performance biases (at least in healthcare where demographic data biases are widespread and pervasive) in a broad class of cases arise due to biases in datasets such as under-representative or misrepresentative training data. Such is the extent of data biases that it makes sense for organizations like the to orient their general bias mitigation advice around data representativeness \citep[c.f.][]{food2021artificial}.}

The explanatory assumption allows for multiple explanatory pathways which chart the way that biases in model performance depend on data biases. One common explanatory pathway that arises in cases of underrepresentative training data is where the model generalizes a statistical association that holds for an overrepresented group to the entire population such that the association fails to hold for the underrepresented group. For example, HbA1c level and stroke are associated in that poor glycemic control is a risk factor for stroke \citep{mitsios2018relationship}. Yet the association between HbA1c level and stroke for women is more pronounced for women over age 55 compared to women under 55 \citep{zhao2014sex}. Hence training data for an HbA1c based stroke risk assessment model that undersamples women over 55 may lead the model to generalize the association that holds for women under 55 to all women, resulting in suboptimal predictive performance for women over 55. Here the underrepresentation of a given demographic group in training data leads the model to underfit for that group. 

Misrepresentative training data can also lead to performance disparities. What is at issue in some instances of misrepresentation is the use of proxy variables that misrepresent a group’s circumstances. For example, \citet{obermeyer2019dissecting} found that a model used in the US to enroll sufficiently sick patients into a narrow care program exhibited racial bias. The model was such that, for any given risk score, Black patients at that risk score were sicker than White patients at that risk score, where sickness is measured by number of active chronic conditions. Patients are automatically referred into the narrow care program only if their risk score is above some threshold. Hence Black patients need to be sicker than White patients to qualify for automatic enrolment in the program. The bias arose because the variable \textit{health cost} was used as a proxy for \textit{health need}. Because Black and White patients matched on health need are on average such that less money is spent on the Black patient, the effect of using health cost as a proxy for health need was to misrepresent the healthcare needs of the Black patients. 

Other cases of misrepresentation involve misrepresentative data labels. This can occur if there are diagnostic disparities across groups. For example, in the United States, clinical screening tools for dementia are known to be less reliable for racial minorities \citep{stephenson2001racial, gianattasio2019racial}.  Models trained on labeled data such that the labels reflect patterns of systematic misdiagnosis are liable to replicate those patterns in their predictions \citep{rajkomar2018ensuring, hellstrom2020bias}. Yet more instances of misrepresentation arise given differential missingness across groups. For example, longitudinal EHR datasets are liable to systematically exclude vulnerable populations such as immigrants who are more likely to receive fractured care across multiple healthcare institutions \citep{gianfrancesco2018potential}. Hence models trained on longitudinal EHR data may fail to detect clinically significant statistical patterns that pertain to the relevant populations. The downstream impact may then be underperformance for certain vulnerable populations. 

The significance of the explanatory assumption relates to bias mitigation. A recent paper by \citet[p.9]{willemink2020preparing} on ‘Preparing Medical Imaging Data for Machine Learning’ states that ‘[i]f an AI algorithm is trained with images from a European institution and the algorithm is used in an Asian population, then performance may be affected by population
or disease prevalence bias.' They continue: ‘It is thus advised to use images from multiple diverse sources, or at least images representing the target population or health system in which the algorithm is to be deployed.’ What is evidenced here is how the explanatory link between algorithmic biases and biases in training data informs algorithmic bias mitigation strategies. In particular, avoiding demographic performance disparities typically requires ensuring that training data appropriately represents the target population.

\section{The Generalist Paradigm}

The main commitment of the narrow paradigm in clinical machine learning is that models are trained for particular clinical use cases such as risk assessment, diagnosis and treatment recommendation. The emerging generalist paradigm differs in that its core focus is on training generalist models that can be adapted for multiple clinical use cases via fine-tuning on narrow biomedical datasets. This section introduces the generalist paradigm and examines in what respects this paradigm challenges the ontic and explanatory assumptions about algorithmic bias in medicine.

\subsection{Self-Supervised Learning}

The models with which we are concerned are large language models such as Open AI’s GPT-2 and GPT-3 \citep{radford2019language, brown2020language}, Google’s BERT and LaMDA \citep{devlin2018bert,thoppilan2022lamda}, and Meta’s OPT \citep{zhang2022opt}. These models are pre-trained for natural language understanding, and can be fine-tuned on domain-specific datasets for particular use cases including clinical use cases. To illustrate: Initially, BERT is pre-trained on English Wikipedia and BooksCorpus for natural language understanding \citep[p.5]{devlin2018bert}. BERT can then be fine-tuned for specific clinical tasks, such as biomedical text mining, through further training on biomedical text datasets such as PubMed \citep[c.f.][]{lee2020biobert}. 

Language models are pre-trained for natural language understanding using a machine learning technique called self-supervised learning \citep[pp.4-5]{bommasani2021opportunities}. Recall that supervised learning involves labeled data. For example, a supervised model for named entity recognition might be trained on ordered pairs containing a biomedical noun and the type of entity to which each noun refers, e.g. (TP53, gene), (thrombin, enzyme). In contrast, self-supervised models are trained on unlabelled text datasets like English Wikipedia \citep{devlin2018bert}. Training tasks on unlabelled text data are automated prediction tasks. For example, one of BERT’s training tasks is called masked language modeling, in which particular words from sentences in the training data are masked such that the model must predict the missing words based on the surrounding words. This task teaches BERT bidirectional context. Given text datasets with the scale and breadth of English Wikipedia, masked language modeling allows language models to learn a broad class of capabilities including arithmetic (‘100 x 20 = \underline{\hspace{0.7cm}}’), general knowledge (‘\underline{\hspace{0.7cm}} is the capital of France), specialist knowledge (‘The $LD_{50}$ of ricin for mice is \underline{\hspace{0.7cm}} mg/kg’), translation (‘\textit{Ospedale} is the Italian word for \underline{\hspace{0.7cm}}’), and unit conversion (‘The adult spleen is approximately 14cm (\underline{\hspace{0.7cm}} inches) in length’). Hence the generalist character of these models.

To be precise: What is learned in the pre-training stage is a mathematical representation of a language called a word embedding. Roughly, a word embedding is a function that maps words to vectors (or points) in a high-dimensional space.\footnote{This formulation is rough because strictly speaking tokens and not words are inputted into the function, i.e. input sequences of text are first tokenized \citep[see][]{grefenstette1999tokenization}.} The geometry of the vector space encodes semantic information in that distances between vectors represent the degree to which words are semantically related. Words with similar meanings map to vectors which are close together. The large language models that we are considering complicate this picture slightly. Notice that certain words have multiple meanings. For example, ‘bank’ can mean the side of a river or it can mean a type of financial institution. Hence it is problematic to map each word to a vector that encodes its meaning, as some words do not have one meaning. Large language models learn what is called a \textit{contextual embedding}, namely, a function that maps each word in an input sequence to a vector in a high-dimensional space, where the vector’s location depends on the entire input sequence as opposed to individual words \citep{Liu2020embeddings}. This allows the model to discriminate between different senses of the same word.

Fine-tuning for clinical use cases differs depending on the pre-trained model’s architecture. It is helpful to distinguish two cases.\footnote{A third case bracketed here is sequence-to-sequence models such as Google’s T5 that include an encoder and a decoder \citep{raffel2020exploring}.} First, \textit{encoder models} such as BERT take text sequences as inputs and output embeddings for each inputted word. Fine-tuning these models for clinical uses typically  involves additional pre-training on biomedical text datasets such as PubMed abstracts and MIMIC clinical notes \citep{peng2019transfer}. The result is an embedding that better registers the semantic nuances of biomedical text data; although in practice whether domain-specific pre-training improves downstream model performance depends on the task at issue \citep{lin2020does}. Biomedical embeddings can then be used as inputs for supervised models that are trained for particular clinical use cases. To illustrate: Suppose the task at issue is hospital readmission prediction using free-text discharge summaries \citep[c.f.][]{huang2019clinicalbert}. Given a labeled dataset comprising discharge summaries plus the ground truth for whether patients were readmitted within the relevant time-window, the idea is to train a supervised model to predict readmission using embeddings for the discharge summaries as inputs.

Second, \textit{decoder models} like GPT-2 \citep{radford2019language} and LaMDA \citep{thoppilan2022lamda} take sequences of text as their inputs and output a prediction for the next word. Feeding the outputs back into the model allows passages of text to be generated. Fine-tuning decoder models for clinical use cases involves next-word prediction tasks on tailored biomedical text datasets. For example, \citet{wang2021evaluation} fine-tuned GPT-2 using transcripts of therapy sessions to evidence the in principle use of language models in therapeutic contexts. Likewise, \citet{sirrianni2022medical} fine-tuned GPT-2 on dental clinical notes to evidence the in principle use of language models for medical charting. In addition to fine-tuning, there is also promising evidence suggesting that pre-trained language models that have not been fine-tuned on domain-specific clinical data encode clinical knowledge \citep{singhal2022large}. The idea is that appropriately tailored prompts can enable the model to accurately answer medical questions.

\subsection{Semantic Biases}

The ontic assumption holds that algorithmic biases in clinical medicine admit analysis in terms of measurable performance disparities across demographic groups. Self-supervised models challenge this conception of algorithmic bias by introducing uniquely semantic biases that are non-reducible to measurable demographic performance disparities. To be clear, semantic biases are understood as a subset of algorithmic biases. What makes these biases semantic is that the nature of these biases has to do with meaning. Consider three examples.\footnote{These examples are illustrative and are not intended as an exhaustive taxonomy.}

Call the first kind of semantic bias \textit{perspectival bias}. What is at issue here is that evaluative commitments may be implicit in the way that words are used or defined. \citet[p.13]{weidinger2021ethical} give an example in which a language model asked ‘What is a family?’ responds ‘A family is a man and a woman who get married and have children.’ Here the meaning attached to the word ‘family’ reflects a heteronormative conception of the family, which is incompatible with conceptions of the family that, for example, make room for families involving same-sex couples or polyamorous familial relationships involving more than two partners.

Perspectival biases are morally significant in clinical use cases for at least two reasons. First, patient-facing conversational interfaces such as therapeutic chatbots or question answering systems may alienate particular patient subpopulations by exhibiting perspectival biases that diminish or deny the significance of certain beliefs, values, identities, or cultural practices. For example, a therapeutic chatbot that indicates a commitment to a hetronormative conception of the family may alienate LGBTQ+ patients and engender the perception that clinical machine learning tools are not developed with LGBTQ+ individuals in mind. What is concerning here is that perspectival biases may be difficult to anticipate in practice and need not involve any explicit diminution of beliefs, values, identities, or practices. For example, an entertainment chatbot on a pediatric ward may ask children whether they are missing home. If pressed to clarify the question, the chatbot may associate ‘home’ with a static residence, which may be alienating for children in the Irish Traveller community or other nomadic communities. Or consider a sexual health chatbot that when pressed to define or elaborate upon terms relevant to gender identity does so using pathologizing or otherwise stigmatizing terminology.

Second, dialogue-based health information systems may exhibit undue influence over the patient’s decision if certain outputs are loaded with perspectival bias. The concern is that a system’s explanation of the costs and benefits of particular treatments may fail to be value neutral in such a way that amounts to the inappropriate distortion of information material to the patient’s decision. This could happen if, for example, qualitative categories reflecting the probability of outcomes (e.g. ‘likely’, ‘unlikely’) or the severity of outcomes (‘severe’, ‘mild’) are used by chatbots to communicate risks of particular interventions \citep{mcneil1982elicitation, halpern2013default}. To illustrate: The claim that Pre-Exposure Prophylaxis (PrEP) for HIV is likely to cause a mild reduction in kidney function may be interpreted by the patient as the claim that the risk can justifiably be ignored. Yet whether the risk can justifiably be ignored depends on the patient’s risk tolerance and the anticipated benefit of PrEP for them given their beliefs and values. Hence the usage of these qualitative risk descriptions may inadvertently impose a particular perspective on the risks of PrEP that inappropriately biases the patient’s decision.

Call the second class of semantic biases \textit{stereotype associations}. To illustrate: In one recent study, \citet{abid2021persistent} found that in 66/100 cases GPT-3 completed the prompt ‘two Muslims walked into a’ with violent phraseology including references to killing and shooting. Relevantly analogous stereotype associations have been observed for LGBTQ individuals, alongside women and people of color \citet{sheng2019woman}. Roughly speaking, the technical explanation for these stereotype biases is that the model infers semantic dependence between, for example, the terms \textit{Muslim} and \textit{terrorism} due to frequent co-occurrence of the terms in the training data \citep{brown2020language}. What is concerning about the possibility of language models exhibiting stereotype associations in conversational interfaces for clinical use cases is the idea that clinical technologies may in some way exacerbate existing stigma and clinical harm faced by marginalized populations.

Call the third kind of semantic bias \textit{semantic haze}. Some discourses can be better represented than others in the internet text datasets such that language models are pre-trained on. It may be the case, for example, that discourse on the United States is more prevalent than discourse on Honduras on particular Reddit datasets, i.e. there is more text about the United States than Honduras on the relevant Reddit pages. That certain discourses feature less prominently as part of text datasets can result in language models having a coarser-grained or ‘hazy’ semantic representation of those discourses. For example, in BERT’s embedding space, the term ‘United States’ is a greater distance from the names of other countries (and thus more semantically distinct) compared to Global South countries which are clustered together and thus less semantically distinct. BERT has a finer-grained semantic representation of discourse pertaining to the United States than to Global South countries \citep[pp.7-8]{zhou2021frequency}.

The ethical significance of semantic haze in clinical use cases is at least threefold. First, there is a clear sense in which patient-facing conversational interfaces that exhibit semantic haze on particular topics may perform less well for some patient groups compared to others. Consider a chatbot that provides sex and relationship information for teenagers. The kinds of questions and concerns that teenagers have around sex and relationships may depend at least in part on cultural circumstances. Indeed, some concerns may require specific cultural knowledge to understand and provide appropriate responses to. Suppose, for example, that a 16-year-old female from a Christian family is engaged in a sexually active relationship with a 15-year old male from an Islamic family. There are concerns within the male’s family around premarital sex, and the family is suggesting to the female to make nikah (the Islamic marriage contract), and potentially to take the \textit{Shahada} (a declaration of belief in God and acceptance of Muhammad as God’s messanger). Semantic haze around Islamic marriage customs given limited exposure to text on Islamic marriage customs in the pre-training dataset may result in suboptimal model performance with respect to providing sex and relationships advice on cases like these. A related issue is that language models may be more likely to hallucinate false information in domains of discourse that are underrepresented in the training data \citep{ji2023survey}. For some clinical use cases, this may result in lower health information quality for certain groups. 

Second, the moral significance of semantic haze is also at least in part accounted for in terms of respect for persons. To illustrate: A known problem in palliative care is that a non-trivial fraction of patients feel ignored, misunderstood, or even silenced by clinicians \citep{norton2003life, frosch2012authoritarian, gramling2016feeling}. This phenomenon is morally significant not only because it causes patients to experience psychological discomfort, but also because failing to listen to and appropriately register patient wishes and concerns plausibly amounts to a failure to afford due respect to patients as persons. There are plausible use cases for language models in palliative care including dialogue agents for companionship \citep{van2021m} and entertainment \citep{garcia2021entertainment}. Here it is at least foreseeable that models exhibiting semantic haze around certain topics may engender a sense of not being heard on the part of patients, such that deployment of models exhibiting the relevant semantic haze may amount to a failure to afford due respect to the patient. Indeed the problem is not restricted to palliative care. Minority patient groups such as transgender patients or patients with disabilities may have a pre-existing sense of being misunderstood or silenced in the healthcare system, and conversational systems exhibiting semantic haze around health topics specific to those groups may invite a sense of their person or their identity having been disrespected among members of the relevant minority groups.

Third, in practice, it may be the case that models which perform poorly or unpredictably for certain groups are not made available to those groups or are subject to certain restrictions on how the models can be used that are not applied to other groups. Consider an online medical question answering tool that reliably answers questions in English but not Spanish. Such a tool may screen prompts with a language classifier so as to detect and refuse prompts written in Spanish. In doing so, the potential benefits of the tool for Spanish speaking users who do not speak English are withheld because the safety risk to that group is intolerably high. Obviously, product decisions like these may compound across multiple tools resulting in patterns of inequity with respect to health information availability across demographic groups. 

\subsection{New Explanatory Pathways}

The last section showed how language models may introduce novel and uniquely semantic biases. These biases matter for conversational use cases such as therapeutic chatbots and question-answering systems. But their significance for non-conversational use cases such as risk assessment or diagnosis based on clinical notes is less apparent. This section argues that in non-conversational use cases language models can complicate the explanatory relationship between biases in training data and familiar model performance biases such as disparities in accuracy across demographic groups. This complication is significant because it destabilizes the standard bias mitigation strategy of ensuring appropriately representative training data. 

Semantic associations learned in pre-training may in some cases partly explain performance disparities in downstream clinical prediction tasks. Consider an example. Suppose that BERT is fine-tuned for the psychiatric use case of EHR-based violence risk assessment \citep{mosteiro2022machine, karystianis2021utilizing}. In pre-training BERT will have learned semantic associations between terms like ‘violence,’ ‘resist,’ and ‘non-compliance,’ and these associations may inform the model’s predictions in the downstream task. The presence of such terms under the right contextual circumstances may be good predictors of violence. Indeed, the principal advantage of BERT-based models for EHR-based violence risk assessment is that BERT embeddings are sensitive to context. Hence the expectation is that semantic associations learned in pre-training affect and improve performance in the downstream clinical task.

However, there is evidence that negative descriptors such as ‘resist’ and ‘non-compliant’ are more likely to feature in EHRs for Black patients compared to non-Black patients \citep{sun2022negative}. Suppose for the purposes of the example that a racial disparity in the use of negative descriptors obtains and is at least partly explained by implicit racial biases on the part of clinicians.\footnote{\citet[p.206]{sun2022negative} note that ‘the use of negative descriptors might not necessarily reflect bias among individual providers; rather, it may reflect a broader systemic acceptability of using negative patient descriptors as a surrogate for identifying structural barriers.’} To clarify: What is being assumed here is that negative descriptors are in fact used more for Black patients compared to non-Black patients and that the explanation for the disparity is that clinicians as a group are disposed at least to some extent to use more negative language when describing interactions with Black patients compared to relevantly similar interactions involving non-Black patients. Given these assumptions, it may be the case that a fine-tuned BERT model for violence risk assessment ends up overpredicting violence for Black patients because the kinds of negative language that best predict violent behavior are disproportionately used to characterize interactions with Black patients. More precisely, the descriptive bias in the EHR data may result in a failure of calibration across racial groups, such that for any given risk score the fraction of actually violent Black patients at that risk score is greater than the fraction of actually violent non-Black patients at that same risk score \citep{hedden2021fairness, kleinberg2016inherent}.

What explains the calibration failure in the above example is misrepresentative training data. Yet the way in which the training data is misrepresentative is novel. What is not at issue is a proxy variable that misrepresents the circumstances of Black patients - as was the case in the \citet{obermeyer2019dissecting} example in which the model used health cost as a proxy for health need. Neither is data missing or mislabelled. Rather the problem is that (by assumption) free-text descriptions of Black patients in EHRs are disproportionately negative in the sense of exaggerating the degree of non-adherence in contrast to non-Black patients who display comparable levels of non-adherence. Language models can register differences in the language used to characterize different demographic groups. Biases encoded in those textual descriptions may translate into model performance biases in a way that is hard to detect. 

Two points are made in closing: (1) It is plausible that clinicians harbor implicit or even explicit biases against disadvantaged groups including racial minorities, people with disabilities, and transgender individuals (see \citet{hall2015implicit} for a review). Such biases may affect clinical notes written about such patients in respects that are not yet well understood and in respects that language models may register. (2) This complicates the standard bias mitigation advice to ensure that clinical training data is representative of the target population.\footnote{Note that this issue is not unique to LLMs. Similar considerations hold for models and research studies that rely on discrete EHR data, which can also encode biases \citep[c.f.][]{ross2020influence}.} For it may be unclear in what respects textual data fails to appropriately represent disadvantaged subpopulations. In particular, where negative leaning language is involved or pathologizing language is used to characterize patient concerns. 

\section{Mitigating Bias}

The emerging generalist paradigm in clinical machine learning complicates the prevailing ontic assumption that algorithmic biases admit analysis in terms of performance disparities across demographic groups. The generalist paradigm also puts question to the prevailing explanatory assumption that biases in model performance mirror biases in training datasets such that bias mitigation strategies ought to focus on demographic representation in biomedical datasets. The upshot is that regulatory policy developed on the backdrop of the narrow paradigm - such as the FDA’s recommendation that ‘[d]ata collection protocols should ensure that the relevant characteristics of the intended patient population [...] are sufficiently represented in [...] test datasets so that results can reasonably be generalized to the population of interest’ - has limited practical significance for clinical developers using generalist models \citep[p.2]{food2021artificial}.

To be sure, biomedical datasets have typically been used to fine-tune generalist models for clinical use cases; and to that end, ensuring appropriate representation within the datasets used for fine-tuning is actionable and relevant advice. Yet two caveats are needed. On one hand, bias mitigation guidance that focuses exclusively on model performance biases that are explained by biases in biomedical datasets occludes considerations relevant to bias mitigation in generalist models. In particular, performance biases in clinical models may be explained by biases in pre-training data which is not specifically biomedical, so there is good reason not to provide bias mitigation guidance which pertains uniquely to biomedical data. On the other hand, there now exist promising examples of clinical information extraction from generalist models that are not fine-tuned on biomedical data \citep{singhal2022large}. Bias guidance that focuses uniquely on the representativeness of biomedical data does not apply to these models and thus excludes an important emerging class of clinical machine learning systems. 

What follows are some plausible directions for bias mitigation guidance in the context of generalist language models applied to clinical use cases. 

\subsection{Pre-Processing and Data Collection}

The first bias mitigation strategy targets the data pre-processing and data collection stages. Pre-processing is an umbrella term for a set of data practices including cleansing datasets for incorrect or missing data and wrangling raw data into an appropriate format for machine learning development. Internet text datasets can be filtered for inappropriate content in the pre-processing stage. The Colossal Clean Crawled Corpus, for example, is a 156 billion token text dataset that has been filtered using the open access ‘Dirty, Naughty, Obscene or Otherwise Bad Words List’ \citep{raffel2020exploring}. The rationale behind filtering text datasets for inappropriate content is that language models are unable to reproduce the kinds of harmful stereotype associations and perspectival biases that arise in, for example, pornographic websites or websites containing far-right political content, if all such websites are removed from the model’s training corpus \citep[p.614]{bender2021dangers}. Hence data filtration is a plausible approach to mitigating semantic biases that may arise in conversational clinical use cases such as therapeutic chatbots.

Nevertheless, filtering text datasets for problematic content involves trade-offs that require articulation and judicious negotiation. On one hand, removing toxic language from training data precludes models from identifying situations in which they are confronted with toxic language and identifying situations in which the model’s own outputs are toxic. Hence data filtration may actually hinder the model’s ability to counteract biases. On the other hand, data filtration can also result in semantic haze around marginalized discourses. \citet[p.614]{bender2021dangers} discuss one example in which the term ‘twink’ features on the ‘Dirty, Naughty, Obscene or Otherwise Bad Words List,’ such that text datasets filtered by this list may systematically exclude LGBTQ+ online spaces. Such exclusion may render the model less able to register semantic distinctions and draw appropriate semantic associations for LGBTQ+ terminology \citep{zhou2021frequency}, which may be problematic in patient-facing conversational use cases such as therapeutic chatbots and question-answering systems. These concerns underscore the importance of active dataset diversification efforts as an important complementary technique alongside filtration for problematic content. 

In addition to these concerns there is a more fundamental worry about who is best placed to decide which content is appropriate for text datasets used to train language models that are subsequently adapted for clinical use cases. These assumptions are potentially problematic in clinical use cases in that patients and clinicians represent a plurality of worldviews and there is an expectation of reasonable disagreement between stakeholders about what counts as appropriate content. While there are plausible mitigations such as accommodating a diversity of cultural backgrounds and political opinions in data filtration decisions, the question of how to establish such an effort in a way that is politically legitimate admits no easy answer. 

\subsection{Fine-Tuning}

The second bias mitigation strategy is to fine-tune the model to avoid harmful or otherwise problematic content. For example, Google’s LaMDA is fine-tuned using supervised learning to ensure appropriate model outputs \citep{thoppilan2022lamda}. In particular, crowdworkers were used to generate conversational material with LaMDA including conversations on sensitive topics and adversarial conversations in which crowd workers attempted to elicit inappropriate model outputs. These outputs were subsequently annotated to identify problematic content, and also the type of problematic content at issue (e.g. religious stereotypes). The resultant dataset was used to fine-tune LaMDA to predict the appropriateness of its outputs which in turn allowed LaMDA to screen-out problematic outputs.\footnote{Ethical fine-tuning can also be achieved via non-supervised approaches such as Reinforcement Learning from Human Feedback \citep{bai2022training,ouyang2022training}. The ethical issues discussed in this section apply to these non-supervised approaches also.} 

Ethical fine-tuning is a promising technique for addressing semantic biases in clinical use cases including stereotype associations and perspectival biases. Certain clinical use cases require the model to \textit{engage} with sensitive content and do so \textit{appropriately}, such as therapeutic use cases in which the model may need to engage in discussions about suicide, drug use, or sex. This gives ethical fine-tuning the advantage over data filtration. Removing a chatbot’s ability to discuss sex by filtering all sexual content from the training data caps the therapeutic utility of the chatbot substantially. In contrast, ethical fine-tuning offers a plausible avenue for chatbots to engage in dialogue around sensitive topics and to do so appropriately. Presumably, mental health professionals as opposed to crowdworkers would be required to construct and label training data involving appropriate and inappropriate therapeutic conversational interactions. This may render ethical fine-tuning for clinical use cases costly given the need for expert clinical labelers. But what matters here is that ethical fine-tuning provides a plausible and scalable mechanism for automated appropriateness checks on clinical model outputs. 

There are, however, at least two limitations to fine-tuning. First, using human annotators to discover failure modes in which the model exhibits harmful biases limits the diversity of failure modes that can be accounted for \citep{perez2022red}. Because clinical models can reasonably be expected to encounter a much broader class of bias-eliciting stimuli in the real-world deployment context, it is at least plausible that fine-tuning based on human-labeled dialogical interactions will lead to critical oversights. The practical significance of this concern depends in large part on the success or failure of emerging strategies for upscaling ethical fine-tuning via automated case generation \citep{perez2022red, ganguli2022red}, and in particular on the applicability of these techniques to conversational clinical use cases. Second, the capacity of generalist language models to exhibit biases is virtually unbounded, especially with respect to perspectival biases which are pervasive in internet text corpora. For that reason it is unclear if the appropriate goal of ethical fine-tuning ought to be perspectival neutrality or rather the adoption of a perspective that is broadly acceptable within the bounds of a pluralistic society. The issue here is that ethical fine-tuning is not itself a solution to the problem of semantic biases in clinical models, but rather a technical method that if supplemented with an appropriate set of evaluative commitments can address the problem. The questions of who is doing the fine-tuning and by what criteria cannot be overlooked. This underscores the need for inclusive and open discussion around how ethical fine-tuning is performed in practice.

\section{Conclusion}

This paper examined how certain pervasive assumptions about the nature and causes of algorithmic bias in medicine are challenged by an emerging class of generalist models such as BERT and GPT-2 that can be fine-tuned for clinical use cases. Specifically, medical regulators such as the FDA in the US and the MHRA in the UK are operating with a conception of bias that is tailored to the dominant narrow paradigm in clinical machine learning, in which supervised models are trained to perform particular tasks such as diagnosis and risk assessment. On this picture, models exhibit biases if and to the extent that they admit measurable performance disparities across demographic groups such as differences in classification accuracy or miscalibrated risk scores. These biases are typically explained by analogous biases in training data such as underrepresentation or misrepresentation of the relevant demographic groups. Generalist models complicate this picture. Not only do these models introduce novel and uniquely semantic biases such as stereotype associations and perspectival biases, they also complicate the explanatory relationship between biases in training data and measurable performance disparities across demographic groups for regression and classification tasks.

Three recommendations are made in closing. First, there is good reason for SaMD medical regulators to explore emerging techniques for bias mitigation in large language models such as data filtration and ethical fine-tuning and to incorporate such techniques into guidance and regulatory policy on the safe and ethical development of clinical machine learning models \citep{bender2021dangers,laurenccon2022bigscience}. Second, healthcare providers seeking to use generalist models for clinical applications ought to consider algorithmic bias not only from the perspective of performance disparities across demographic groups relative to regression and classification metrics, but also consider auditing models for clinically significant semantic biases. Third, medical ethicists have good reason to examine the moral significance of biases in clinical models beyond considerations pertaining to the equitable distribution of benefits and burdens in healthcare. For example, examining the class of respect-based moral complaints that patients may have towards models exhibiting biases that reflect harmful stereotype associations.

\section*{Acknowledgements}

I am grateful to Michael Howell, Heather Cole-Lewis, Lisa Lehmann, Diane Korngiebel, Thomas Douglas, Bakul Patel, Kate Weber, and Rachel Gruner for helpful comments, alongside participants at the Workshop on the Ethics of Influence at the Uehiro Centre for Practical Ethics at the University of Oxford.

\bibliographystyle{plain}
\bibliography{references}

%\nocite{*}

\end{document}